\documentstyle{elsart}
\input amssym.def
\newcommand{\beq}{\begin{equation}}
\newcommand{\eeq}{\end{equation}}
\newcommand{\barr}{\begin{array}}
\newcommand{\earr}{\end{array}}
\newcommand{\ord}{\mathop{\rm ord}}
\newcommand{\tr}{\mathop{\rm tr}}
\newcommand{\sgn}{\mathop{\rm sgn}}
\renewcommand{\Re}{\mathop{\rm Re}}
\renewcommand{\Im}{\mathop{\rm Im}}
\newtheorem{theorem}{Theorem}
\newtheorem{lemma}{Lemma}

\newcounter{one}
\newcounter{two}
\begin{document}

\begin{frontmatter}
\title{Approximation theorem for the self-focusing Nonlinear Schr\"odinger
Equation and for the periodic curves in ${\bf R}^3$.}
\author{P.G.Grinevich\thanksref{G1}}
\thanks[G1]{This work was supported by the INTAS grant No. 96-770
and by the Russian Foundation for Basic Research grant No. 98-01-01161.}
\address{Landau Institute for Theoretical Physics,
Kosygina 2, Moscow, 117940, Russia. \\ e-mail: pgg@landau.ac.ru}
\vskip 1em
 \rightline{\small\it Dedicated to V.E.Zakharov's 60th
birthday}
\begin{abstract}
It is shown, that any sufficiently smooth periodic solution of the
self-focusing Nonlinear Schr\"odinger equation can be approximated
by finite-gap ones with an arbitrary small error. As a corollary
an analogous result for the motion of closed curves in ${\Bbb R}^3$
guided by the Filament equation is proved. This equation describes
the dynamics of very thin filament vortices in a fluid.
\end{abstract}
\end{frontmatter}

One of the basic questions of the finite-gap theory is the following:
are the finite-gap solutions of a given equation sufficiently generic
or they belong to some special subclass? To answer this question it
is reasonable to check if arbitrary periodic in spatial variables
solution can be approximated by finite-gap ones.

The study of simplest examples shows, that the uniform approximation
for all $x$ and $t$ is impossible, because to do it we have to keep
all space and time frequencies simultaneously, and we have too
many constraints on the deformations to fulfill all of them.

Therefore it is natural to ask the following questions:

1) Let us have a smooth periodic in spatial variables solution
and an arbitrary compact domain $U$ in the $(x,t)$ space. Is it possible
to approximate our solution on $U$ by finite-gap ones with arbitrary
small error? The approximating solutions are allowed to be non-periodic.
We shall call such approximations {\bf local}.

2) Let us have a smooth periodic in spatial variables solution
and an arbitrary compact domain $U$ in the $t$ space. Is it possible
to approximate our solution by periodic in $x$ finite-gap ones with
the same $x$-periods for all $x$ and $t\in U$?
We shall call such approximations {\bf periodic}.

Of course any periodic approximation is automatically local, but
the characterization of periodic finite-gap solutions is usually
sufficiently complicated, therefore the transition from local
approximations to periodic ones may be rather nontrivial.

In our text we construct {\bf periodic} approximations for the
self-focusing Nonlinear Schr\"odinger equation (SfNLS)
\beq
\label{nls_1}
iq_t+q_{xx}+2q^2\bar q=0, \ \ \ \mbox{where}
\eeq
$q=q(x,t)$ is a complex-valued function of two real variables,
and for the Filament equation
\beq
\label{filament_1}
\frac{\partial\vec\gamma(s,t)}{\partial t}=k(x,t)\vec b(s,t),
\ \ \ \mbox{where}
\eeq
$\vec\gamma(s,t)$ is a $t$-dependent family of smooth curves in ${\Bbb R}^3$,
$s$ is the natural parameter (i.e. $|\partial_s\vec\gamma(s,t)|\equiv1$),
$\vec b(s,t)$, $k(s,t)$ denote the binormal vector from the Frenet reper
and the curvature function respectively.

The first periodic approximation theorem was proved in 1975 by
Marchenko and Ostrovskii \cite{MO75} for the real Korteveg de Vries
(KdV) equation. The method of \cite{MO75} is based on the theory of
conformal maps and it can be naturally extended to some
soliton systems including the defocusing NLS. But for
other systems the question is still open and the answer depends on the
equation. For example, from results of Krichever
\cite{Kr93} it is rather clear that any periodic
Kadomthsev-Petviashvili~\Roman{two} (KP~\Roman{two}) solution allows
finite-gap approximations, but it is likely that  for KP~\Roman{one} it
is not so.

Direct attempts to generalize the approach of \cite{MO75} to SfNLS meet
the following problems
\begin{enumerate}
\item The space of spectral curves corresponding to real periodic $g$-gap
KdV solutions is topologically ${\Bbb R}\oplus\left({\Bbb R}^+\right)^g$.
But in the SfNLS case this space it the real part of some ramified covering
of ${\Bbb C}^{g+1}$, and the structure of the Marchenko-Ostrovskii conformal
map is essentially more complicated. To avoid a detailed study of the
parameters space we use the method of isoperiodic deformations suggested
by M.Schmidt and author in \cite{GS95}.

\item In the KdV case the characterization of admissible divisors is very
simple: the $n$-th point of divisor is located at an arbitrary point of the
$n$-th compact real oval. In the SfNLS case the characterization is less
explicit (see below) and we have to describe how we vary the divisor after
perturbing the spectral curve to preserve the admissibility.

\item Dubrovin equations for the real KdV are non-singular therefore a
small change of parameters and starting point slightly affects the solution.
But in the SfNLS case Dubrovin equations may have singularities, and the
solutions may have branch points (nevertheless the corresponding SfNLS
potential is smooth). Therefore we have to check that our variations
of spectral data do not change the solution too much. To do it we
introduce some new ``symmetric'' variables.
\end{enumerate}

Let us recall some basic facts from the SfNLS theory. The
scattering transform for NLS was found in 1971 by Zakharov and
Shabat \cite{ZS71}. Finite-gap NLS solutions were first
constructed in 1976 by Its and Kotljarov \cite{Its-Kotl76}. The
characterization of SfNLS admissible divisors as well as a proof
that all solutions with reduction (\ref{lax4})  are automatically
nonsingular were obtained by Cherednik \cite{Ch80}. Infinite-gap
periodic problem for matrix operators including the NLS
$L$-operator was studied by M.Schmidt \cite{Sch96}. A lot of
useful information about the NLS theory including the Hamiltonian
theory is contained in the book \cite{FT80} by Faddeev and
Takhtadjan. Finite-gap NLS theory is discussed in details in the
article \cite{Pre85} by Previato. Effictivisation of low genus
formulas by NLS was studied by Kamchatnov \cite{Kam90}.

The zero-curvature representation for SfNLS reads as:
\beq
\label{lax1}
\frac{\partial \Psi(\lambda,x,t)}{\partial x}=U(\lambda,x,t)
\Psi(\lambda,x,t), \ \
\frac{\partial \Psi(\lambda,x,t)}{\partial t}=V(\lambda,x,t)
\Psi(\lambda,x,t),
\eeq
where $\Psi(\lambda,x,t)$ is a 2-component vector,
\beq
\label{lax2}
\Psi(\lambda,x,t)=\left[\begin{array}{c} \psi_1(\lambda,x,t)\\
\psi_2(\lambda,x,t) \end{array}
\right],
\eeq
$U(\lambda,x,t)$,  $V(\lambda,x,t)$
are the following $2\times2$ matrices:
\beq
\label{lax3}
U(\lambda,x,t)=
\left[\begin{array}{cc}i\lambda & iq(x,t) \\ i r(x,t) & -i\lambda
\end{array}\right],
V(\lambda,x,t)=-2\lambda U(\lambda,x,t)+
\left[\begin{array}{cc}iqr & -q_x \\ r_x & -iqr  \end{array}\right],
\eeq
\beq
\label{lax4}
r(x,t)=\bar q(x,t).
\eeq
We shall assume that $q(x,t)$ is periodic in $x$ with the period 1

\beq
\label{per1}
q(x+1,t)\equiv q(x,t).
\eeq

We shall fix our attention on the spectral transform for a fixed $t=t_0$,
therefore starting from this moment we shall omit $t$ in our notations.

The Bloch eigenfunction $\Psi(\lambda,x)$ is by definition the common
eigenfunction of $L=\partial_x-U(\lambda,x)$ and the shift operator

\beq
\label{bloch1}
L\Psi(\lambda,x)=0, \ \ \ \Psi(\lambda,x+1)=e^{ip(\lambda)}\Psi(\lambda,x).
\eeq

Equation (\ref{bloch1}) defines $\Psi(\lambda,x)$ up to a constant factor.
We fix it assuming

\beq
\label{bloch2}
\Phi(\lambda,0)\equiv1, \ \ \mbox{where} \ \ \
\Phi(\lambda,x)=\psi_1(\lambda,x)+\psi_2(\lambda,x).
\eeq

The function $p(\lambda)$ is defined up to adding $2\pi n$, $n\in {\Bbb Z}$.
It is called the quasimomentum.

To calculate the Bloch function we have to diagonalize the $2\times2$
monodromy matrix $T(\lambda)$, which is an entire function of $\lambda$.
The eigenfunctions of $T(\lambda)$ lie on a two-sheeted covering $\Gamma$
of the $\lambda$-plane. $\Gamma$ is called spectral curve. Denote the
permutation of sheets of $\Gamma$ by $\sigma$. $\det T(\lambda)\equiv1$,
therefore $p(\gamma)+p(\sigma\gamma)\equiv 0(\mbox{mod}2\pi)$.
(We shall denote points of $\Gamma$ by $\gamma$ and the projection
$\Gamma\rightarrow{\Bbb C}$ by ${\cal P}$, $\lambda={\cal P}\gamma$).
$p(\gamma)$ is a locally holomorphic multivalued function
on $\Gamma$, $dp=(d p(\lambda)/d\lambda)d\lambda$ is a holomorphic
differential on the finite part of $\Gamma$, $\sigma(dp)=-dp$.
$p(\gamma)\sim\pm\lambda$ as $\lambda\rightarrow\infty$
(as an asymptotic series), therefore $\Gamma$ is compactified by
2 infinite points $\infty_+$, $\infty_-$, $\sigma\infty_+=\infty_-$,
$\sigma\infty_-=\infty_+$, $p(\gamma)\sim\pm\lambda$ as
$\gamma\rightarrow\infty_\pm$ respectively.

A point $\lambda\in{\Bbb C}$ is called {\bf regular} if
$p(\gamma)\ne0(\mbox{mod}\pi)$, where ${\cal P}\gamma=\lambda$ and
{\bf irregular} otherwise. Let $\lambda_k$ be an irregular point.
The Tailor expansion of $\tr T(\lambda)$ reads as:
$\tr T(\lambda)=\pm2+T_k^n(\lambda-\lambda_k)^n+\ldots$. Let us call
$n$ the order of the point $\lambda_k$, $n=\ord_q(\lambda_k)$.
($q$ means that the order is defined in terms of the quasimomentum function).
$\lambda_k$ is a {\bf branch point} of $\Gamma$ if $n$ is odd and a
{\bf double point} of $\Gamma$ if $n$ is even. A branch point is called
{\bf simple} if $n=1$. If the opposite  is not stated explicitly we have
{\bf one} point of $\Gamma$ over each double point. To have an uniform
representation for our equations we shall treat an irregular point of
order $n$ as the result of fusion $n$ simple branch point.

If $\Psi(\lambda,x)$ is a Bloch solution of (\ref{bloch1}), then
\beq
\label{conj1}
\Psi^+(\bar\lambda,x)=
\left[\begin{array}{c} \bar\psi_2(\lambda,x)\\-\bar\psi_1(\lambda,x)
\end{array} \right],
\eeq
is also a Bloch solution of (\ref{bloch1}) with the quasimomentum
$p^+(\bar\lambda)=-\bar p(\lambda)$. Therefore $\Gamma$ has the following
antiholomorphic involutions $\gamma\rightarrow\sigma\bar\gamma$ and
$\gamma\rightarrow\bar\gamma$.
(We assume that $p(\bar\gamma)=\bar p(\gamma)$,  $p(\sigma\bar\gamma)=
-\bar p(\gamma)$). For real $\lambda$ $T(\lambda)$ is an unitary matrix,
$p(\lambda)\in{\Bbb R}$, and $\Gamma$ has no real branch points
(but may have real double points).

Let $\{E_k\}$,  $\{E^+_k\}$ be the lists of all irregular points, the
index $k$ takes all integer values. We assume that
\begin{enumerate}
\item $\Im E_k\ge 0$.
\item $E_k^+=\overline{E_k}$.
\item If $\ord_q(\lambda_k)=n$ the point $\lambda_k$ has exactly $n$
entries in our lists. For example if $\ord_q(\lambda_k)=4$ and
$\lambda_k\in{\Bbb R}$ then we have exactly 2 integers $k_1$, $k_2$
such that $E_{k_1}=E_{k_2}=E^+_{k_1}=E^+_{k_2}$.
\end{enumerate}

\begin{lemma}
It is possible to enumerate the irregular points so, that for
sufficiently large $|k|$
\begin{enumerate}
\item $E_k=(\pi\sgn k )\cdot\sqrt{k^2-I_1(q)}+o\left(\frac{1}{k}\right)$
where $I_1(q)=\int_0^1q(x)\bar q(x)dx$.
\item $E_k-E^+_k\rightarrow0$ faster than any degree of $k^{-1}$ as
$|k|\rightarrow\infty$ (We assume $q(x)$ to be smooth).
\end{enumerate}
\end{lemma}

The next important object for us is the set of {\bf zeroes of the
quasimomentum differential} $dp$. They are invariant under the involution
$\sigma$ therefore we shall consider their projections to the
$\lambda$-plane instead. Denote them by $\alpha_k$
where $k$ takes all integer values. As above we use the following agreement
\begin{enumerate}
\item If $\lambda_k$ is a regular points it has $n$ entries in the list
$\{\alpha_k\}$ where $n$ is the order of zero of $dp$ at one sheet.
\item If $\lambda_k$ is an irregular points it has
$\left[\frac{\ord_q(\lambda_k)}{2}\right]$ entries to the list
$\{\alpha_k\}$ where $[\ ]$ denotes the integer part.
\end{enumerate}

\begin{lemma}
It is possible to enumerate the points $\alpha_k$ so, that for sufficiently
large $|k|$
\begin{enumerate}
\item $\Im \alpha_k=0$.
\item $\alpha_k=\Re E_k+o(\Im E_k)$.
\end{enumerate}
\end{lemma}

Let us define now the ``second part'' of the spectral data --
the {\bf divisor of poles} of the Bloch function.

Let $\tilde\Psi(\gamma,x)$ denote a Bloch eigenfunction of $L$ with some
non-singular locally holomorphic normalisation
(of course $\Psi(\gamma,x)=\tilde\Psi(\gamma,x)/\tilde\Phi(\gamma,0)$
where $\tilde\Phi(\gamma,x)=\tilde\psi_1(\gamma,x)+\tilde\psi_2(\gamma,x)$.
Consider the Wronskian of the Bloch functions
$\tilde W(\gamma)=\tilde\psi_1(\gamma,x)\tilde\psi_2(\sigma\gamma,x)-
\tilde\psi_2(\gamma,x)\tilde\psi_1(\sigma\gamma,x)$. It is defined up to a
non-zero holomorphic multiplier and does not wanish at regular points.
Let $\lambda_k$ be an irregular point. We have $W(\lambda)=
\pm w_k^m(\lambda-\lambda_k)^{m/2}(1+o(1))$ where
$m$ is even if $\lambda_k$ is a double point and odd if
$\lambda_k$ is a branch point, $m\ge0$.
Denote $m$ by $\ord_b(\lambda_k)$. It is easy to check that
$\ord_b(\lambda_k)\le \ord_q(\lambda_k)$.
A double point $\lambda_k$ is called {\bf removable} if
$\ord_b(\lambda_k)=0$. It is well-known that removable double points
can be treated as regular points and we can forget about them.

{\bf The divisor of Bloch function zeroes} is a list of points of $\Gamma$
$\{\gamma_k(x)\}$ where $k$ takes all integer values such that each zero of
$\tilde\Phi(\gamma,x)$ generates $l$ entries to this list if
$l$ is the multiplicity of it and each irregular point generates
$(\ord_q-\ord_b)/2$ entries. {\bf The divisor of Bloch function poles}
$\{\gamma_k\}$ coinsides with the divisor of Bloch function zeroes taken
at the point $x=0$.

\begin{lemma}
\begin{enumerate}
\item The spectral curve $\Gamma$ has only finite number of non-removable
double points and degenerate branch points.
\item All real double points are removable.
\end{enumerate}
\end{lemma}

\begin{lemma}
It is possible to enumerate the points $\gamma_k$ so, that for sufficiently
large $|k|$ ${\cal P}\gamma_k=\Re E_k+O(\Im E_k)$.
\end{lemma}

It is well-known, that the spectral curve and the divisor of poles
completely define the potential $q(x)$. To reconstruct the potential we
can use {\bf Dubrovin equations}
\beq
\label{dubr1}
\frac{\partial}{\partial x}\lambda_j(x)=-2i\left[\lambda_j(x)+
\sum\limits_{k=-\infty}^{\infty}
\left(\frac{E_k+E^+_k}{2}-\lambda_k(x) \right) \right]\nu_j(x),
\eeq
where
\beq
\label{dubr2}
\nu_j(x)=\sqrt{(\lambda_j(x)-E_j)(\lambda_j(x)-E^+_j)}
\prod\limits_{k\ne j}\frac{\sqrt{(\lambda_j(x)-E_k)(\lambda_j(x)-E^+_k)}}
{\lambda_j(x)-\lambda_k(x)},
\eeq
and the reconstruction formula
\beq
\label{dubr3}
q(x)=\sum\limits_{k=-\infty}^{\infty}
\left(\frac{E_k+E^+_k}{2}-\lambda_k(x) \right)+
\sum\limits_{j=-\infty}^{\infty}\nu_j(x).
\eeq
The infinite sums and products in the formulas above perfectly converge.

Let us recall the characterization of  divisors corresponding to
operators with the reduction (\ref{lax4}). Consider the following 1-form
on $\Gamma$: $\Omega(\gamma,x)= \omega(\gamma,x) d\lambda$, where
\beq
\label{omega1}
\omega(\gamma,x) =
\frac
{(\tilde\psi_1(\gamma,x)+\tilde\psi_2(\gamma,x))
(\tilde\psi_2(\sigma\gamma,x)-\tilde\psi_1(\sigma\gamma,x))}
{\tilde\psi_1(\gamma,x)\tilde\psi_2(\sigma\gamma,x)-
\tilde\psi_2(\gamma,x)\tilde\psi_1(\sigma\gamma,x)}.
\eeq
Denote by  $U(R)$ be the domain $|\lambda|<R$ in $\Gamma$. Consider the
following function $\omega_R(\gamma,x)= \omega(\gamma,x)
\prod\limits_{k : |E_k|<R}\sqrt{(\lambda-E_k) (\lambda-E^+_k)}$.
Denote by $D(\omega_R,x)$ the divisor of zeroes of $\omega_R(\gamma,x)$
and by $D(\omega,x)$  the limit of $D(\omega_R,x)$ as $R\rightarrow\infty$.

\begin{lemma}
\label{l-direct}
$D(\omega,x)$ coincide with the set  $\{\gamma_k(x),\bar\gamma_k(x)\}$,
where $\{\gamma_k(x)\}$ is the divisor of Bloch function zeroes.
\end{lemma}

Let $\delta_j$ be an arbitratry collection of pairwise distinct real points
such that for sufficiently large $|j|$ $\delta_j =\Re E_j$.

\begin{lemma}
\label{l-omega}
\begin{enumerate}
\item
The form $\Omega$ reads as $\Omega=\left[1-\tilde\kappa(\gamma,x)\right]
d\lambda$
\beq
\label{omega2}
\tilde\kappa(\gamma,x)=\sum\limits_{j=-\infty}^{\infty}\frac{\kappa_j(x)}
{\sqrt{(\lambda-E_j)(\lambda-E^+_j)}} \prod\limits_{k\ne j}
\frac{\lambda-\delta_k}{\sqrt{(\lambda-E_k)(\lambda-E^+_k)}},
\eeq
where $\kappa_j(x)$ are some real functions of $x$.
\item $|\kappa(\gamma,x)|\le1$ for all  $x\in{\Bbb R}$, $\gamma\in{\Bbb R}$.
\end{enumerate}
\end{lemma}

In particular $\kappa(\delta_j,x )|\le1$ for all $j$. It gives us the
following estimate on the functions $\kappa_j(x)$:
\beq
\label{omega3}
|\kappa_j(x)|\le \left|\sqrt{(\delta_j-E_j)(\delta_j-E^+_j)}
\prod\limits_{k\ne j} \frac{\sqrt{(\delta_j-E_k)(\delta_j-E^+_k)}}
{\delta_j-\delta_k}\right|\le \left| \delta_j-E_j \right| C_\Gamma,
\eeq
where $ C_\Gamma$ is a positive constant, depending only on the spectral
curve. In particular, if we have a removable double point
$E_k=E^+_k=\delta_k$, then $\kappa_k(x)\equiv0$.

\begin{lemma}
\label{l-inverse}
Let $\kappa_k(0)$ be a collection of real numbers such, that
$|\kappa(\gamma,0)|\le1$ for all $\gamma\in\Gamma$, where
$\kappa(\gamma,0)$ is defined by (\ref{omega2}), $D(\omega,0)$ be the
corresponding divisor, $\{\gamma_j(0)\}$ be any set of points
such that $ D(\omega,0)=\{\gamma_k(0),\bar\gamma_k(0)\}$. Then the
corresponding operator $L$-operator satisfy (\ref{lax4}), and the potential
$q(x)$ is nonsingular.
\end{lemma}

For us the following definition will be convenient: potential $q(x)$ is
called {\bf finite-gap} if $E_j=E^+_j$ for all $|j|\ge J_0$. Then all
points $E_j=E^+_j$ are removable double points,  $\alpha_j=\gamma_j(x)=E_j$
for all $|j|\ge J_0$ and $\Gamma$ has only finite number of branch points
and non-removable double points. Finite-gap solutions of soliton equations
were first introduced by Novikov in 1974 for KdV \cite{Nov74}. The
corresponding solutions can be written explicitly in terms of Riemann
$\theta$-functions.

The first step of the approximation procedure is to construct a finite-gap
deformation of $\Gamma$ generating solutions with the same period. To do it
we need the following lemma proved by M.Schmidt and the author in \cite{GS95}.

\begin{lemma}
Let $\alpha_k\in{\Bbb R}$ be the projection of  a zero of the quasimomentum
such, that $\alpha_k\ne\alpha_j$ for $j\ne k$, $\alpha_k\ne E_j$,
$\alpha_k\ne E^+_j$, for all $j$. Consider the following system of ODE's,
associated with the point $\alpha_k$:
\begin{displaymath}
\frac{\partial E_j}{\partial \tau}=-\frac{c_k(\tau)}{E_j-\alpha_k}, \ \ \
\frac{\partial E^+_j}{\partial \tau}=-\frac{c_k(\tau)}{E^+_j-\alpha_k},
\end{displaymath}
\beq
\label{isodef1}
\frac{\partial \alpha_j}{\partial \tau}=-\frac{c_k(\tau)}{\alpha_j-\alpha_k}
\ \ \mbox{ for } \ \  j \ne k,
\eeq
\begin{displaymath}
\frac{\partial \alpha_k}{\partial \tau}= c_k(\tau)
\left[ \sum_{j\ne k}\frac{1}{\alpha_j-\alpha_k}-
\frac12\sum_{j=-\infty}^{\infty}
\left( \frac{1}{E_j-\alpha_k}
+\frac{1}{E^+_j-\alpha_k}\right)\right],
\end{displaymath}
where $c_k(\tau)$ is an arbitrary real function of $\tau$.

Denote by $\Gamma(\tau)$ the solution of (\ref{isodef1}) with the initial
value $\Gamma(0)= \Gamma$, where the spectral curve $\Gamma$ corresponds to
a periodic with the period 1 potential $q(x)$ (of course this solution is
defined only in some neighborhood of zero $U(0)$). Then for all
$\tau\in U(0)$ the curve $\Gamma(\tau)$ generates periodic with the
period 1 potentials (the $x$-quasifrequencies of the potentials do not
depend on the divisor).
\end{lemma}

Let $|k|$ be sufficiently large. Then using this deformation we can merge
the pair $E_k$, $E^+_k$ to a removable double point, and the corresponding
shift of all points $E_j$, $E^+_j$, $\alpha_j$ with $j\ne k$ is of order
$o(\Im E_k)$. Therefore applying this deformation to all $k$ such that
$|k|\ge K$, where $K>0$ is a sufficiently large integer, we obtain a
finite-gap spectral curve $\Gamma_K$ (it is almost evident that the
superposition of infinitely many deformations perfectly converges).
\begin{lemma}
For any $\epsilon>0$ there exists a $K$ such, that
\begin{enumerate}
\item $|E_j-\tilde E_j|<\epsilon$, $|E^+_j-\tilde E^+_j|<\epsilon$,
$|\alpha_j-\tilde\alpha_j|<\epsilon$, for all $j$.
\item $|\Im(E_j-\tilde E_j)|<\epsilon |\Im E_j| $, for all $j$ such,
that  $|j|<K$.
\end{enumerate}
where $\tilde E_j$, $\tilde E^+_j$, $\tilde\alpha_j$ are the branch points
of the curve $\Gamma_K$ and the quasimomentum zeroes respectively.
\end{lemma}

We have constructed a family of finite-gap curves $\Gamma_K$ approximating
the curve $\Gamma$. Let us discuss now the admissible divisors.

\begin{lemma}
There exists a pair of positive integer constants $K_1$, $K_2$ such, that
for all $K\ge K_2$ the points of any admissible divisor $\gamma_k$ on
$\Gamma_K$ can be enumerated so, that
\begin{enumerate}
\item For all $k$  such that $|k|\le K_1$ $|\lambda_k|<K_1+1/10$.
\item For all $k$ such that $|k|> K_1$
$|(\lambda_k-\tilde\delta_k)|\le\Im\tilde E_k$, and
$|\tilde\delta_k|>K_1+1-1/10$.
\end{enumerate}
\end{lemma}

The proof  follows from the characterization of admissible divisors given by
Lemmas~\ref{l-direct}-\ref{l-inverse}.

Equations (\ref{dubr1})-(\ref{dubr2}) have singularities at the right-hand
side. To simplify the structure of Dubrovin equation it is convenient
to introduce the following new variables:
\begin{enumerate}
\item $s_k(x)$, $q_k(x)$, $1\le k\le 2K_1+1$ -- the first $2K_1+1$ expansion
coefficients at $\infty$ of the function
\beq
\label{kappa1}
\Xi(\gamma,x)=\left( 1+\tilde\kappa(\gamma,x)\right) \prod_{k}
\frac{\sqrt{(\lambda-E_k) (\lambda-E^+_k)}}{(\lambda-\lambda_k(x))}
\eeq
\beq
\label{kappa2}
\Xi(\lambda,x)= \pm\left(1+\sum_{k>0} \frac{s_k(x)}{\lambda^k} \right)+
\sum_{k>0} \frac{q_k(x)}{\lambda^k}  \ \ \mbox {as} \ \
\gamma\rightarrow\pm\infty.
\eeq
\item $\tilde\lambda_k(x)=\lambda_k(x)-\delta_k$, $|k|>K_1$.
\item $\tilde\nu_k(x)= \sqrt{(\lambda_k(x)-E_k) (\lambda_k(x)-E^+_k)}$,
$|k|>K_1$.
\end{enumerate}

These variables are dependent. Denote this set of variables by ${\cal S}$.

Consider the following norm:
\beq
\label{norm}
\|  {\cal S} \|_n=\sqrt{\sum_{|k|\le K_1}\left( |s_k|^2+|q_k|^2 \right) +
\sum_{|k|> K_1}|k|^n \left( |\tilde\lambda_k|^2+|\tilde\nu_k|^2 \right)}
\eeq

This norm is bounded on the space of admissible divisors for any positive
$n$.

\begin{lemma}
For any sufficiently large $n$ there exists a constant $C_n(\Gamma)$ such,
that for any admissible pair ${\cal S}_1(x)$, ${\cal S}_2(x)$ of
solutions of Dubrovin equations we have the following estimate
\beq
\label{est1}
\left \| \frac{\partial}{\partial x}\left({\cal S}_1(x)-
{\cal S}_2(x)\right )\right \|_n
\le C_n(\Gamma) \| {\cal S}_1(x)-{\cal S}_2(x) \|_n.
\eeq
\end{lemma}

It is easy to check, that for any $\epsilon_1>0$ there exists a
constant $K_3(n)$ such that
\beq
\label{est2}
\| {\cal S} \|_n^{(2)}<\epsilon_1, \ \ \mbox{where} \ \ \
\| {\cal S} \|_n^{(2)} = \sqrt{\sum_{|k|> K_3(n)}|k|^n
\left( |\tilde\lambda_k|^2+|\tilde\nu_k|^2 \right)}.
\eeq

We need also the following semi-norm
\beq
\label{norm2}
\|  {\cal S} \|_n^{(1)}=\sqrt{\sum_{|k|\le K_1}
\left(|s_k|^2+|q_k|^2 \right) +
\sum_{K_1<|k|< K_3(n)}|k|^n \left( |\tilde\lambda_k|^2+
|\tilde\nu_k|^2 \right)}.
\eeq

It is evident, that $\|  {\cal S} \|_n^\le \|  {\cal S} \|_n^{(1)}+\|
{\cal S} \|_n^{(2)}$ and $\|  {\cal S} \|_n^{(1)}\le\ \| {\cal S} \|_n$.

Combining (\ref{est1}) and (\ref{est2}) we obtain the following estimate
\beq
\label{est3}
\| {\cal S}_1(x)-{\cal S}_2(x) \|_n\le e^{C_n(\Gamma)|x|}
\left (\| {\cal S}_1(0)-{\cal S}_2(0) \|_n^{(1)}+\epsilon_1\right) .
\eeq

Therefore in approximate  calculations we can truncate the Dubrovin system
to a finite-dimensional one, removing the variables $\tilde\lambda_k(x)$,
$\tilde\nu_k(x)$, with the $|k|>K_3(n)$.  For the truncated system
small variations of the curve and of the starting point result in small
variations of the solution. To complete the proof it is sufficient to
check, that choosing $K$ sufficiently large we can make the admissible
variation of divisor arbitrary small. But it follows from the
characterization of admissible divisors presented above.

These arguments can be applied also for the Dubrovin equations, describing
the $t_l$ -evolution of the divisor, where $t_l$ denotes the $l$-s time
from the NLS hierarchy  (in these notations $t=t_1$). Taking into account
that the first $k$ $x$-derivatives of $q(x)$ are continuous functionals
in the norm $\| \ \|_n$ for sufficiently large $n$ we obtain:

\begin{theorem}
Let $q(x,t)$ be an arbitrary SfNLS solution with smooth $x$-periodic Cauchy
data $q(x,0)=q_0(x)$. Then for any $\epsilon>0$, $N>0$ and ${\cal T}>0$
there exists a finite-gap  SfNLS solution $q^F(x,t)$ such, that
\beq
\left| \frac{\partial^n}{\partial x^n} \left( q^F(x,t)-q(x,t) \right)
\right |<\epsilon
\ \ \ \mbox {for all} \ \ \ x\in{\Bbb R}, \ |t|<{\cal T}, \  n\le N.
\eeq
\end{theorem}

At the end let us say a few words how to prove an analogous theorem for the
Filament equation.
Equivalence between the SfNLS and the Filament equation is given by the
Hasimoto map \cite{Has72}. ($\theta$-functional solutions of Filament
equations were studied by Sym in \cite{Sym}).
\beq
\label{hasimoto}
q(s,t)=\frac{1}{2} k(s,t)e^{i\int^s \kappa(\tilde s,t)d\tilde s},
\eeq
where $k(s,t)$, $\kappa(s,t)$ are the curvature and the torsion functions
respectively. In \cite{GS97} it was shown, that equations (\ref{isodef1})
(except one corresponding to $\alpha_0=0$)  preserve the periodicity in
$s$ of the Filament equation solution. Therefore the technique developed
above can be applied without changes.

{\bf Acknowledgments.} The author is grateful to Martin Schmidt for
explaining some results from the periodic NLS theory.

\end{document}